\documentclass[prc,showpacs,amsmath,amssymb,floatfix]{revtex4}

\usepackage{graphicx}
\usepackage{epsfig}
\usepackage{slashed}
\usepackage{dcolumn}

\begin{document}

\title{Dilepton production in pion--nucleon collisions in an effective
  field theory approach}

\author{Mikl\'os Z\'et\'enyi}
\email{zetenyi.miklos@wigner.mta.hu}
\author{Gy\"orgy Wolf}
\email{wolf.gyorgy@wigner.mta.hu}
\affiliation{Institute for Particle and Nuclear Physics,
Wigner Research Centre for Physics, Hungarian Academy of Sciences, P.O. Box 49, H-1525 Budapest, Hungary}

\date{\today}

\begin{abstract}
  We present a model of electron-positron pair production in pion-nucleon
  collisions in the exclusive reaction $\pi N \rightarrow Ne^+e^-$. The
  model is based on an effective field theory approach, incorporating 16
  baryon resonances below 2 GeV. Parameters of the model are fitted to
  pion photoproduction data.  We present the resulting dilepton invariant
  mass spectra for $\pi^- p$ collisions up to $\sqrt{s}=1.9$~GeV
  center-of-mass collision energy. These results are meant to give
  predictions for the planned experiments at the HADES spectrometer in
  GSI, Darmstadt.
\end{abstract}


\maketitle

\section{Introduction}

Dileptons are among the most important signals studied in heavy ion
collision experiments. In the 1--2 GeV/nucleon energy range
electron-positron pair production has been studied by the DiLepton
Spectrometer (DLS) at LBL and, more recently, by the High Acceptance
Di-Electron Spectrometer (HADES) at GSI. Due to the high complexity of
nuclear collision processes, the experimental results can be interpreted
only via comparison with model calculations. Usually transport models
are used for this purpose. These models need the cross sections of
elementary hadronic collisions as input, therefore a good understanding
of the elementary cross sections is essential.

Both DLS \cite{DLS_NN} and HADES \cite{HADES_NN} studied dilepton
production in elementary $NN$ collisions. In parallel a lot of
theoretical work has been done in order to achieve a good description of the
experimental dilepton spectrum. Earlier, a resonance approach was used
\cite{Wolf_dilep,Bratkovskaya}, where particle production is described
as a multistep process. In the first step a baryon resonance is created
which then decays in one or more steps, creating the final state
particles, including dileptons. This approach naturally fits the
particle production mechanism of transport codes. Recent calculations
apply one-boson-exchange effective Lagrangians to calculate the $NN
\rightarrow NNe^+e^-$ cross section
\cite{Shyam2003,Kaptari2006,Kaptari2007,Shyam2009,Kaptari2009,Shyam2010}.
Although a lot of progress has been made, the measured dilepton spectra
are still not perfectly reproduced by the theoretical models
\cite{Shyam2010}.

In heavy ion collisions a large number of pions are produced, therefore
elementary $\pi N$ collisions are also important. Moreover, besides
photon induced reactions, pion beams are much more suitable for studying
individual resonances than nuclear projectiles.  At HADES new
experiments are planned with a pion beam, where both $\pi A$ and $\pi N$
collisions would be studied. At the same time, dilepton production in
$\pi N$ collisions have not yet been studied in an effective field
theory approach similar to those used in the $NN$ case.

The process $\pi N \rightarrow Ne^+e^-$ is related to the time inverse
of pion photoproduction, which is the key experiment in determining the
electromagnetic properties of baryon resonances, and is studied in great
detail both experimentally and theoretically. In particular, effective
field theory models have been used to study pion photoproduction
\cite{Garcilazo,Feuster_Mosel_NPA,Fernandez}.

In the present paper we set up a model of electron-positron pair
production in $\pi N$ collisions based on an effective field theory
approach.

The paper is organized as follows. In Sec.~\ref{sec:kinema} we review
the kinematics of the $\pi N \rightarrow Ne^+e^-$ process and give the
expressions for the differential cross section. In Sec.~\ref{sec:EFT} we
specify the effective Lagrangians and discuss the calculation of the
transition matrix elements. Separate subsections deal with the version
of the vector meson dominance model used in this paper to describe the
electromagnetic interaction of hadrons; the contribution of the
nonresonant Feynman diagrams to the matrix element, with an emphasis on
the gauge-invariance preserving scheme for hadronic form factors; the
contribution of baryon resonances. For nonresonant contributions
explicit analytical expressions for the matrix elements are listed,
while the contributions of baryon resonances are calculated numerically.

In Sec.~\ref{sec:resparam} we discuss the determination of baryon
resonance parameters from pion photoproduction data.  The calculated
dilepton spectra are shown in Sec.~\ref{sec:results}, followed by a
discussion.

\section{\label{sec:kinema} Kinematics}
\begin{figure}[htb]
  \begin{center}
    \includegraphics[width=4cm]{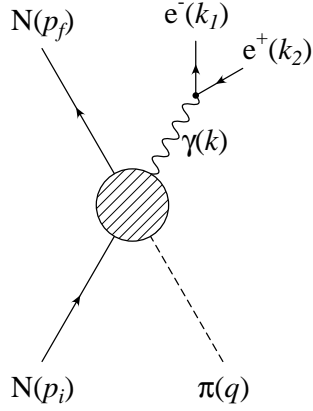}
    \caption{\label{fig:kinema} Schematic diagram of the process $\pi +
      N \rightarrow N + e^+ + e^-$. }
  \end{center}
\end{figure}

The differential cross section of the process $\pi + N \rightarrow N +
e^+ + e^-$ is given by
\begin{equation}
  \label{eq:dsigma}
  d\sigma = \frac{(2\pi)^4}{4\sqrt{(p_i \cdot q)^2 - m_N^2 m_{\pi}^2}}
  \frac{1}{n_\text{pol}} \sum_\text{pol} \left\vert\mathcal{M}\right\vert^2
  d\Phi_3\left( p_i+q; p_f,k_1,k_2\right),
\end{equation}
where the $n$-body phase-space is defined by
\begin{equation}
  \label{eq:phase-space}
  d\Phi_n\left( P; p_1,...,p_n\right) =
  \delta^{(4)}\left(P-\sum_{i=1}^{n}p_i\right) \prod_{i=1}^{n} \frac{d^3
    \mathbf{p}_i}{(2\pi)^3 2p_{i0}},
\end{equation}
and we have used the notation of Fig.~\ref{fig:kinema} for the
four-momenta of particles. The three-body phase-space in
Eq.~(\ref{eq:dsigma}) can be calculated recursively as
\begin{equation}
  d\Phi_3\left( p_i+q; p_f,k_1,k_2\right) = (2\pi)^3 d(k^2)
  d\Phi_2\left( p_i+q; p_f,k\right) d\Phi_2\left( k; k_1,k_2\right).
\end{equation}
Making use of the Dirac-$\delta$ in Eq.~(\ref{eq:phase-space}) we can
integrate out four of the six momentum components in the case of the
two-body phase-space, to get
\begin{equation}
  \label{eq:phase-space2}
  d\Phi_2\left( P; p_1,p_2\right) =
  \frac{1}{4(2\pi)^6}\frac{|\mathbf{p}_1|}{\sqrt{P^2}} d\Omega_1,
\end{equation}
where $\mathbf{p}_1$ is the spatial part of $p_1$ in the frame where $P$
is at rest, and $d\Omega_1 = d\phi_1 d(\cos\theta_1)$ is the solid angle
of $p_1$ in the same reference frame.

Using this the differential cross section Eq.~(\ref{eq:dsigma}) can be
written in the form
\begin{equation}
  d\sigma = \frac{1}{64(2\pi)^5|\mathbf{q}|s} d(k^2) d\Omega_\mathbf{k}
  d\Omega_{\mathbf{k}_1} \frac{|\mathbf{k}||\mathbf{k}_1|}{\sqrt{k^2}}
  \frac{1}{n_\text{pol}} \sum_\text{pol}
  \left\vert\mathcal{M}\right\vert^2.
\end{equation}
For unpolarized beams $d\sigma$ is independent of the azimuth angle
$\phi_\mathbf{k}$, which can be integrated out.  Note that
$\mathbf{k}_1$ and $d\Omega_{\mathbf{k}_1}$ is defined in the rest frame
of the decaying virtual photon of momentum $k$, in accordance with
Eq.~(\ref{eq:phase-space2}). Further, $\sqrt{k^2} = M$ is the dilepton
invariant mass, and $d(k^2) = d(M^2) = 2MdM$. Neglecting the electron
mass we get $|\mathbf{k}_1| = M/2$. The differential cross section is
then
\begin{equation}
  \label{eq:dsdm}
  \frac{d\sigma}{dM} = \frac{M}{64(2\pi)^4 s}
  \frac{|\mathbf{k}|}{|\mathbf{q}|} \int d(\cos\theta_\mathbf{k})
  d\Omega_{\mathbf{k}_1} \frac{1}{n_\text{pol}} \sum_\text{pol}
  \left\vert\mathcal{M}\right\vert^2.
\end{equation}
In Eq.~(\ref{eq:dsdm}) the magnitudes of the center-of-mass momenta are
given by
\begin{eqnarray}
  |\mathbf{q}| & = & \frac{\sqrt{\lambda(s,m_N^2,m_\pi^2)}}{2\sqrt{s}} \\
  |\mathbf{k}| & = & \frac{\sqrt{\lambda(s,m_N^2,M^2)}}{2\sqrt{s}},
\end{eqnarray}
with $\lambda(a,b,c) = a^2 + b^2 + c^2 - 2(ab + bc + ca)$.

The leptonic part of the matrix element $\mathcal{M}$ can be written out
explicitly, resulting in the expression
\begin{equation}
  \label{eq:matrixelement}
  \mathcal{M} = - \frac{e}{k^2}\mathcal{M}_{\mu}^{\text{had}}
  \bar{u}(k_1)\gamma^{\mu}v(k_2).
\end{equation}
The squared matrix element summed over polarizations is
\begin{equation}
  \sum_\text{pol} \left\vert\mathcal{M}\right\vert^2 = \frac{e^2}{k^4}
  W_{\mu\nu}l^{\mu\nu},
\end{equation}
with the hadronic tensor $W_{\mu\nu}$ defined by
\begin{equation}
  W_{\mu\nu} = \sum_\text{pol}
  \mathcal{M}_{\mu}^{\text{had}}{\mathcal{M}_{\nu}^{\text{had}}}^{*},
\end{equation}
and the leptonic tensor $l^{\mu\nu}$ given by
\begin{equation}
  l^{\mu\nu} = 4\left(k_1^{\mu}k_2^{\nu} + k_1^{\nu}k_2^{\mu} -
  (k_1\cdot k_2)g^{\mu\nu} \right).
\end{equation}

\section{\label{sec:EFT} Effective Lagrangians and Matrix Elements}

\begin{figure}
  \begin{center}
    \includegraphics[width=3cm]{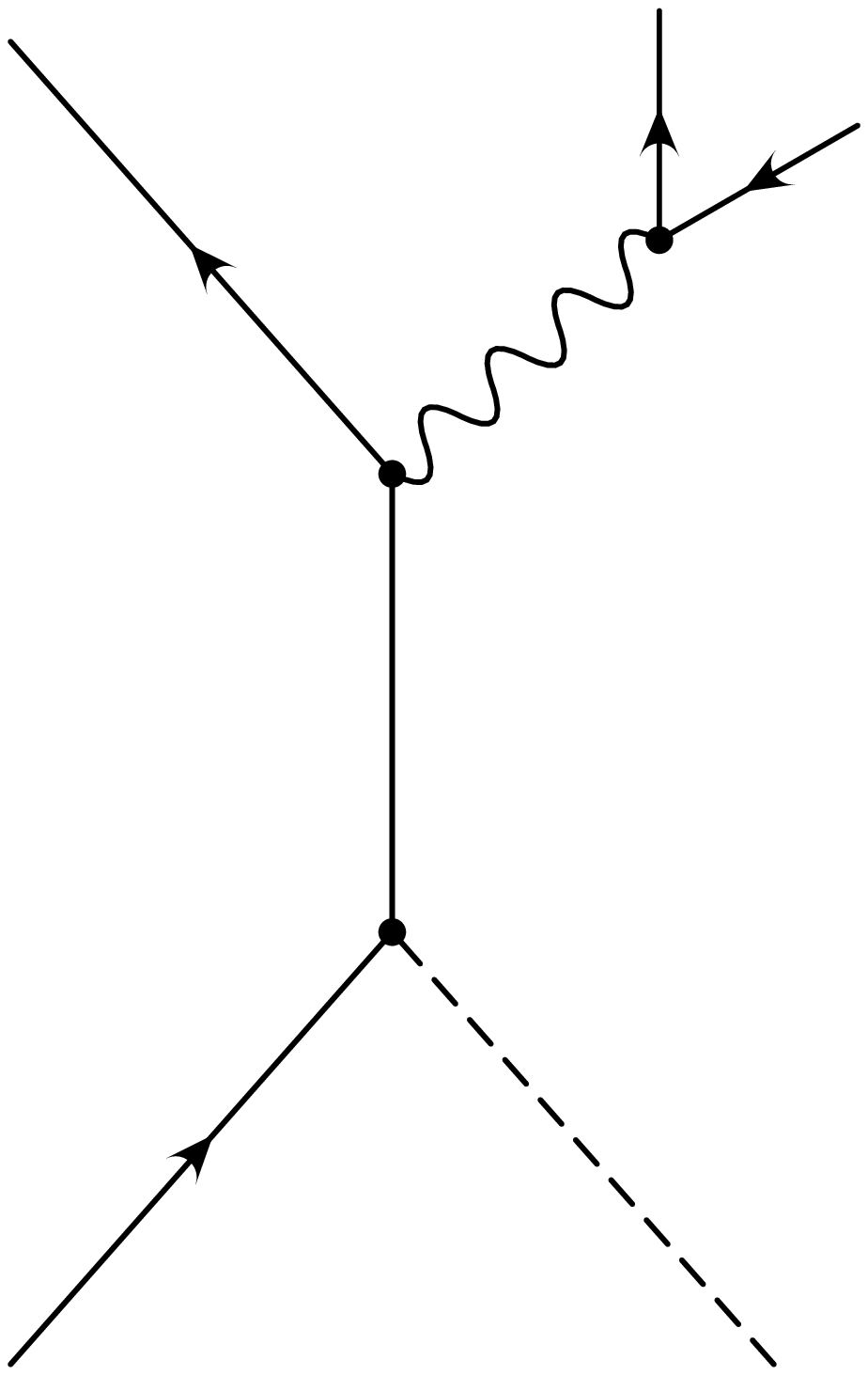} \hspace{0.5cm}
    \includegraphics[width=3cm]{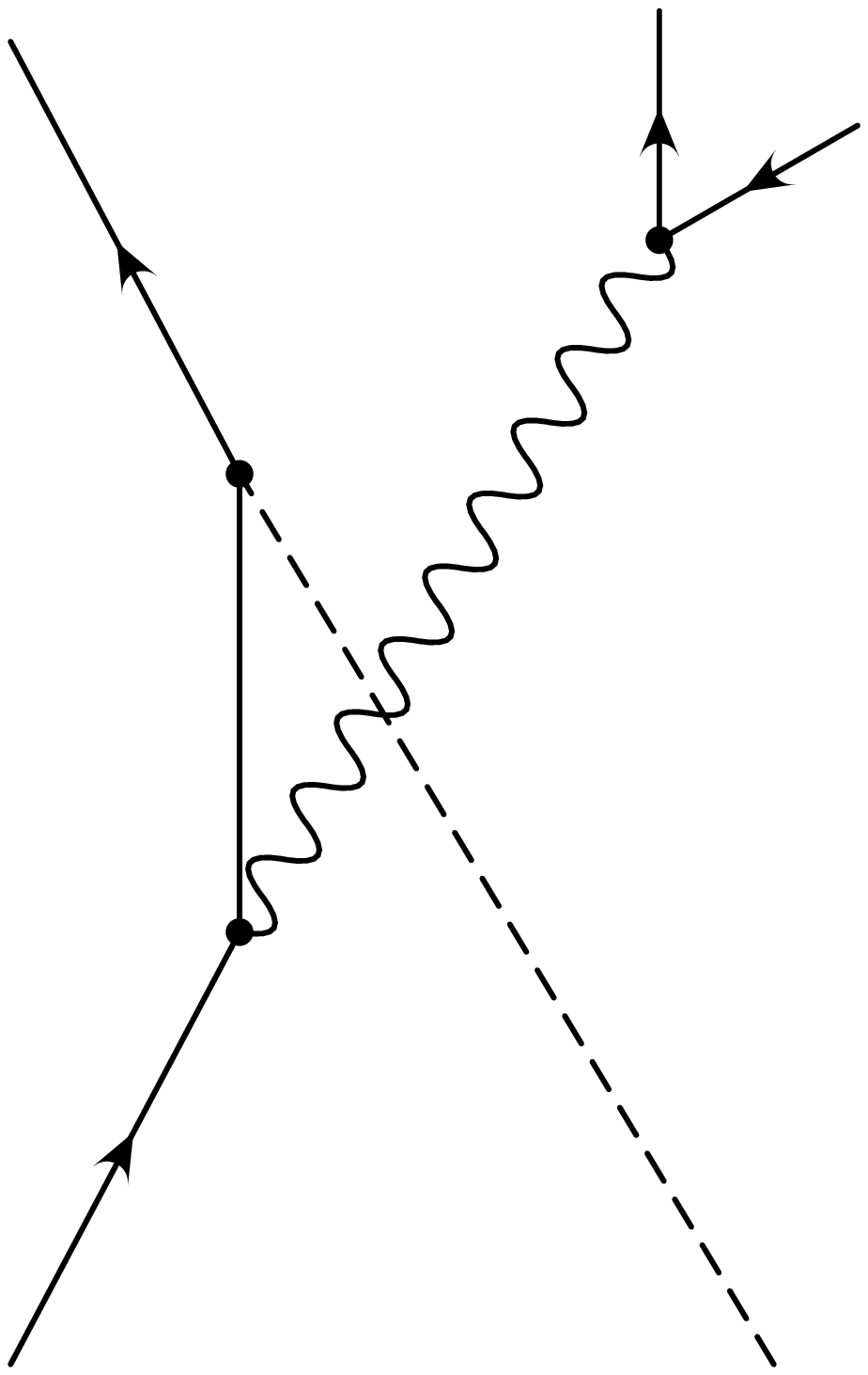} \hspace{0.5cm}
    \includegraphics[width=3cm]{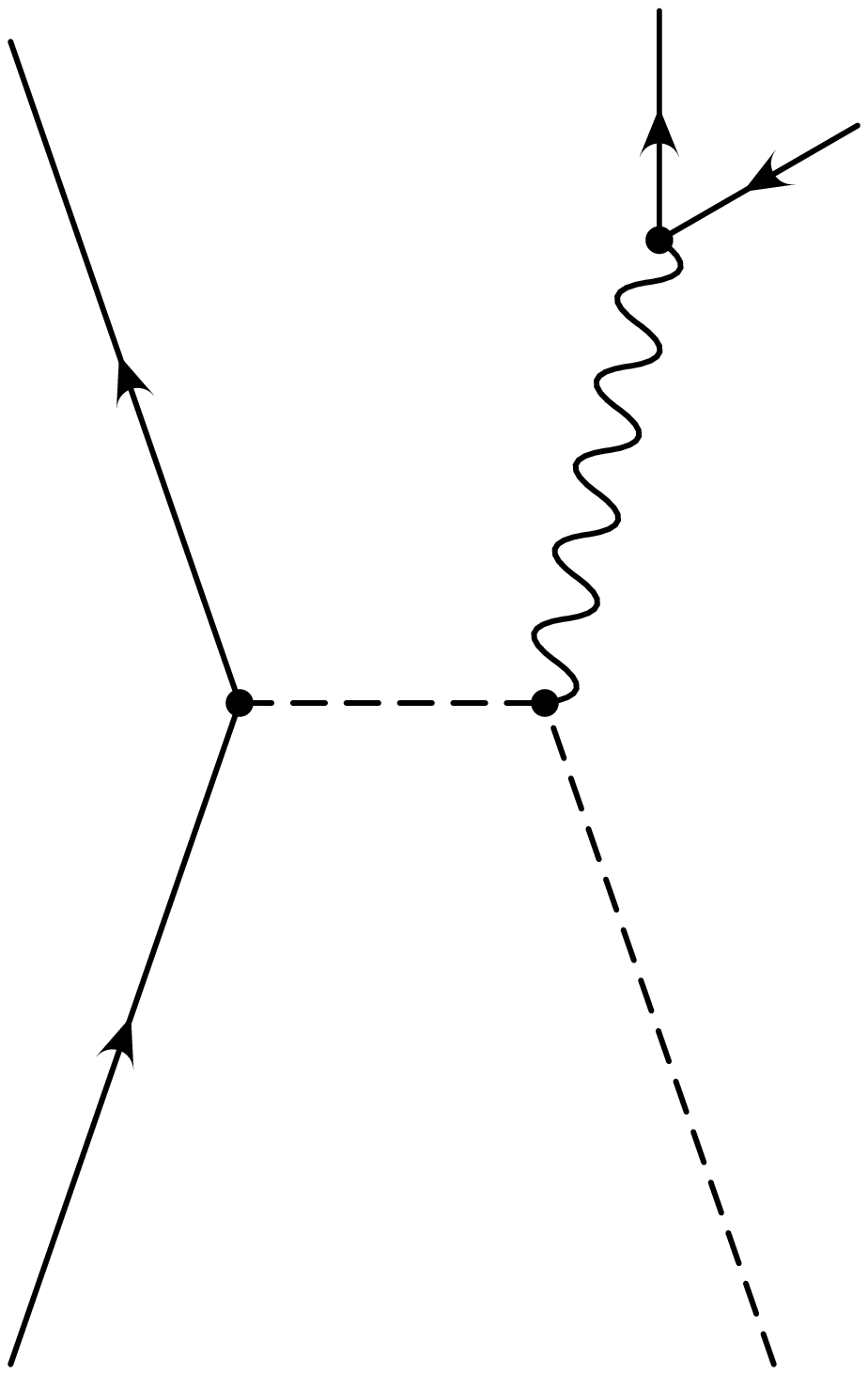} \hspace{0.5cm}
    \includegraphics[width=3cm]{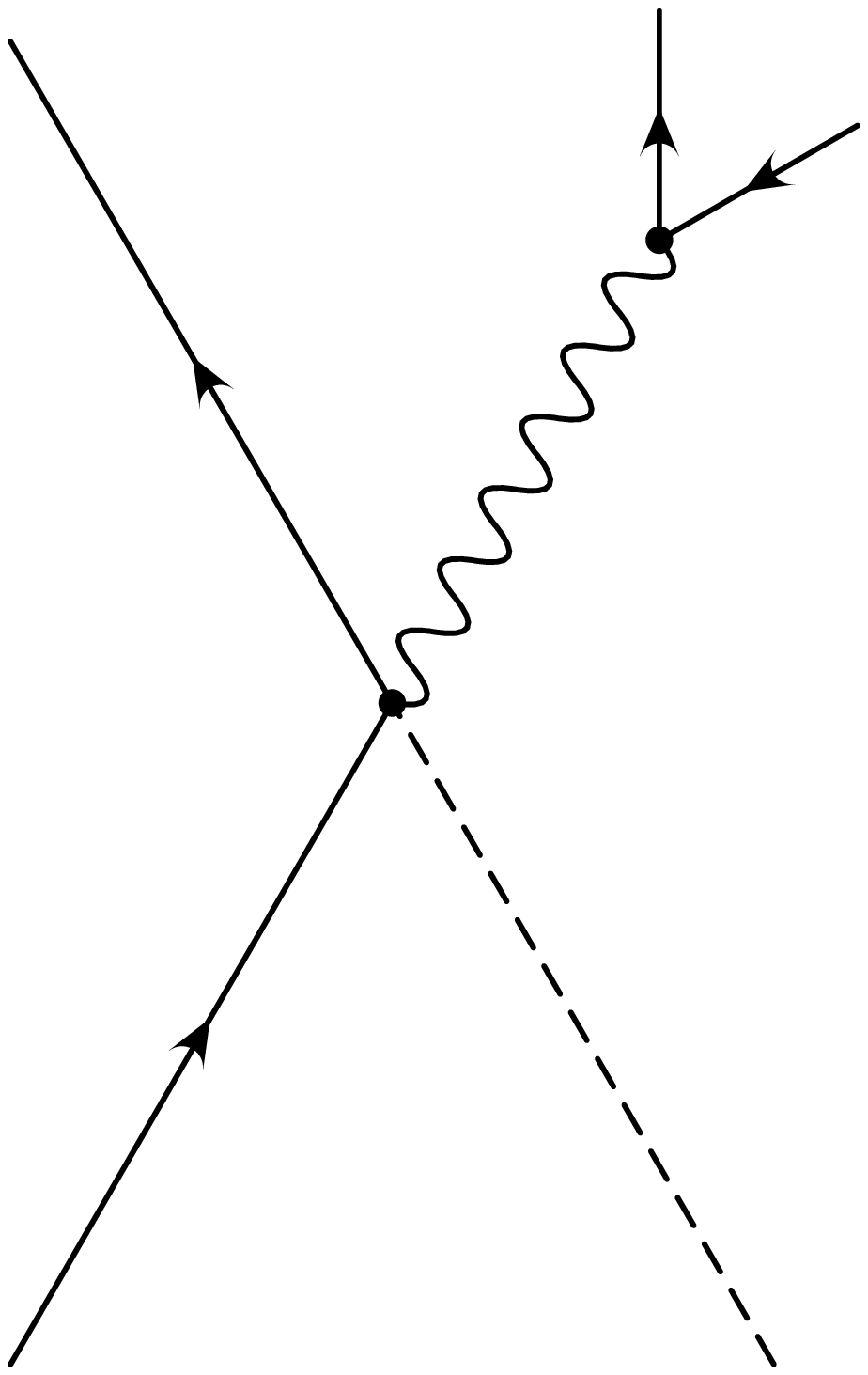} \\ \vspace{0.5cm}
    \includegraphics[width=3cm]{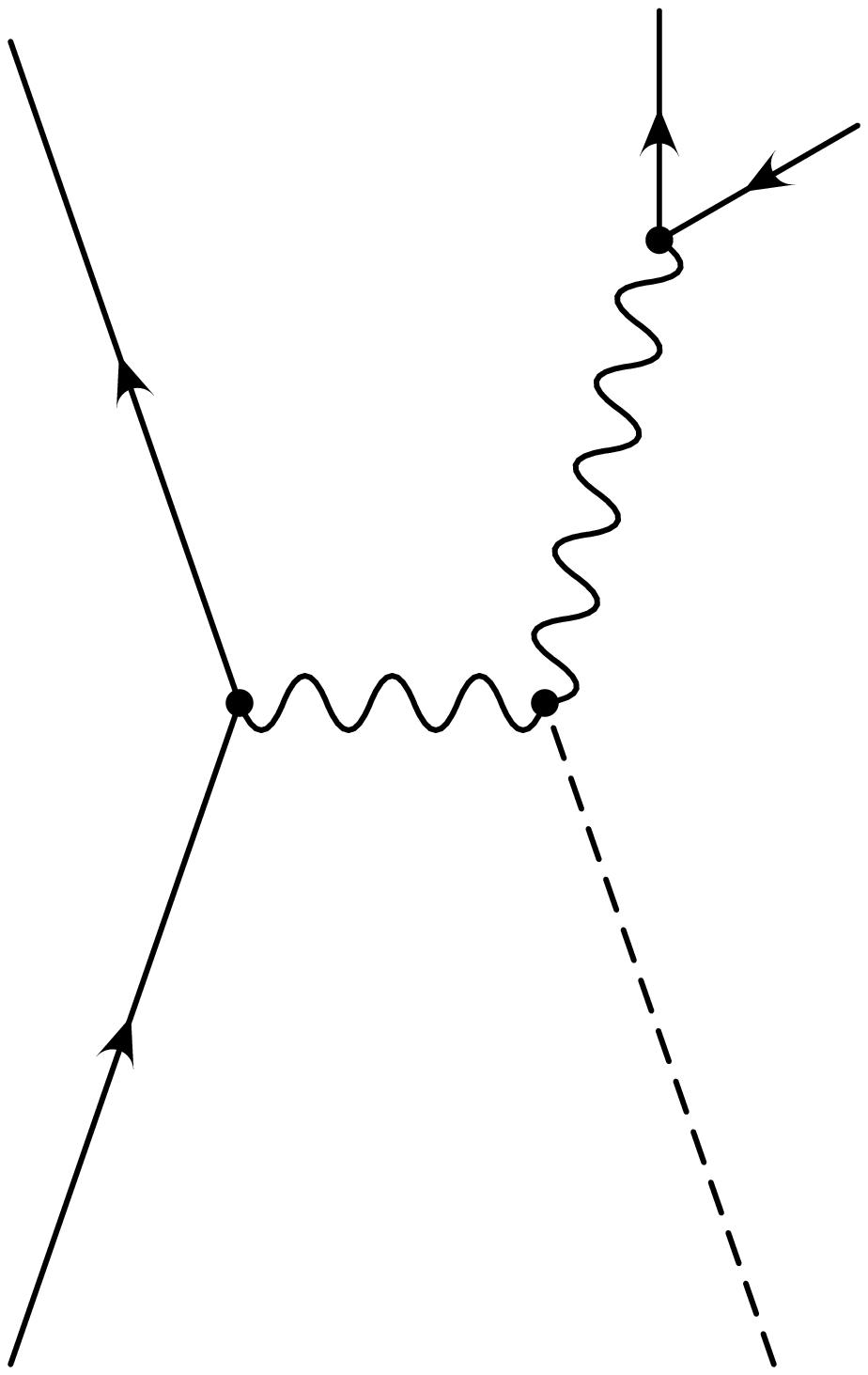} \hspace{0.5cm}
    \includegraphics[width=3cm]{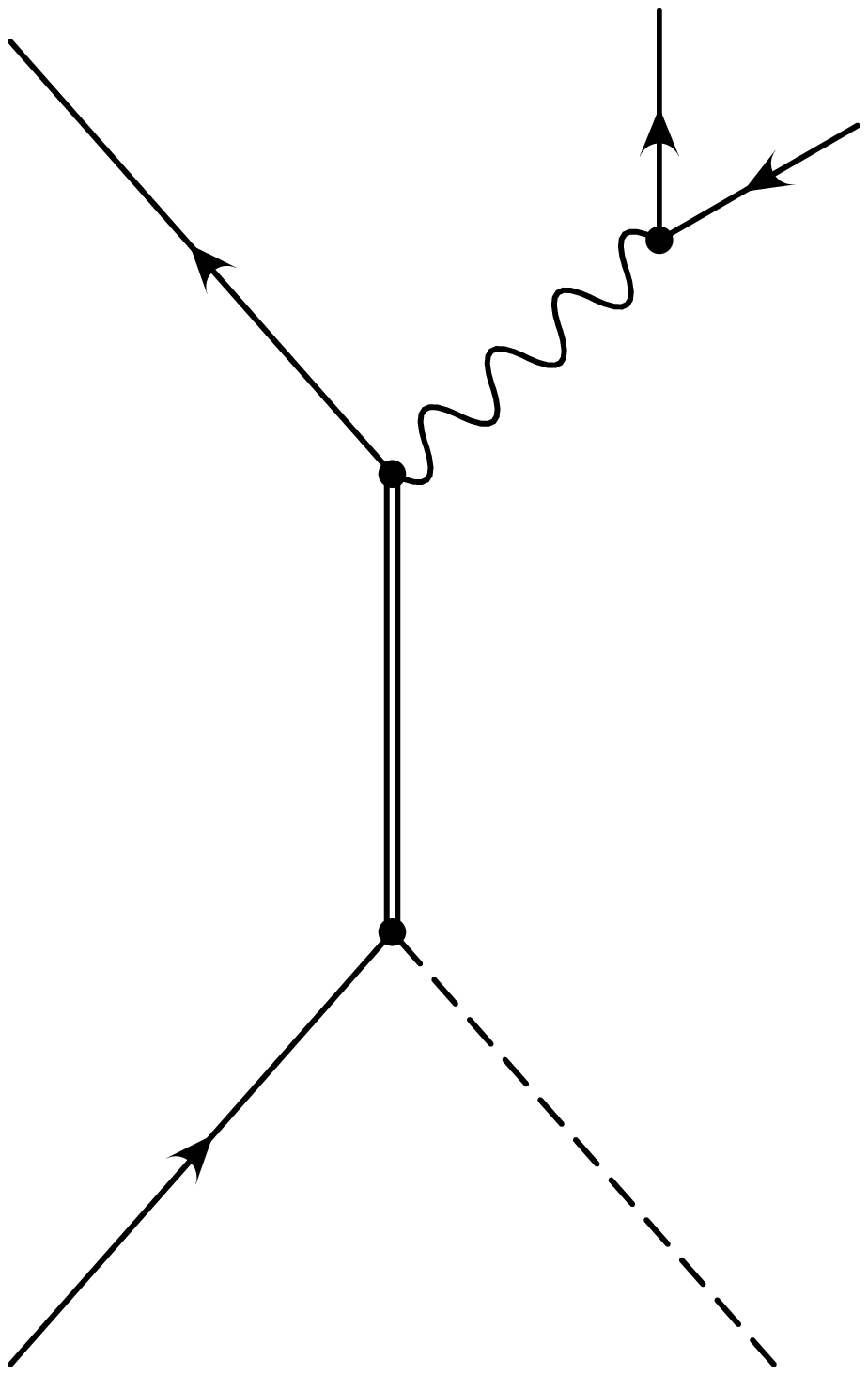} \hspace{0.5cm}
    \includegraphics[width=3cm]{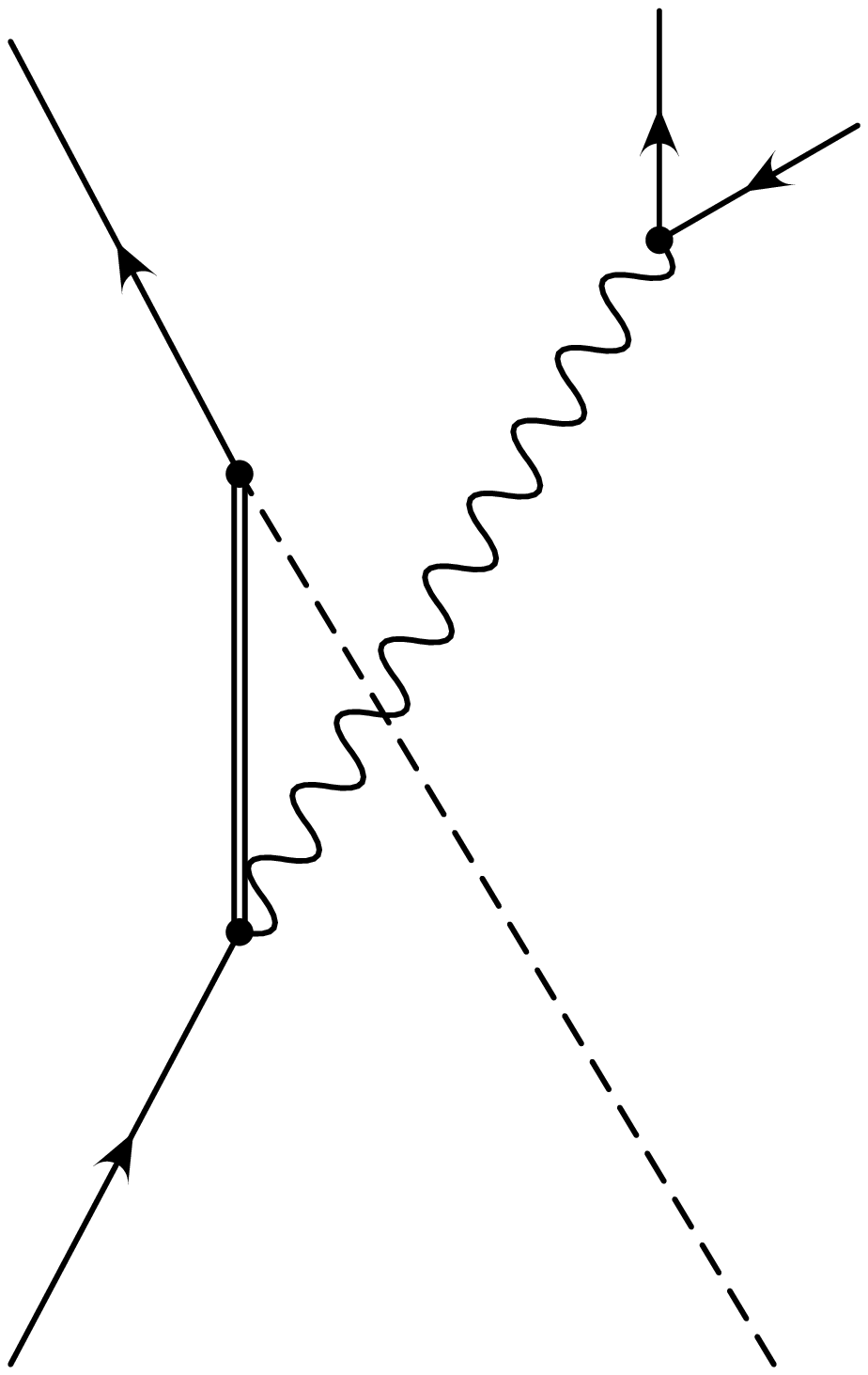} \hspace{0.5cm}
    \caption{\label{fig:diagrams} Feynman diagrams contributing to the
      process $\pi + N \rightarrow N + e^+ + e^-$.}
  \end{center}
\end{figure}

The Feynman diagrams contributing to the process $\pi + N \rightarrow N
+ e^+ + e^-$ are depicted in Fig.~\ref{fig:diagrams}. These are: the
Born contributions [(a) $s$-, (b) $u$-, and (c) $t$-channel diagrams, and
(d) contact interaction term], (e) vector meson exchange diagram, (f)
$s$-channel and (g) $u$-channel baryon resonance contributions.

\subsection{Electromagnetic interaction of hadrons} 
In most studies the electromagnetic interaction of hadrons is described
using some variant of the vector meson dominance (VMD) model \cite{VMD}.
Here we adopt a version of the model described in Appendix B of
Ref.~\cite{Kroll} and also in Ref.~\cite{VMD1}, where it is denoted
VMD1. In this version only the $\rho^0$ vector meson is included and the
$\rho\gamma$ coupling has the form
\begin{equation}
  \label{eq:VMD}
  \mathcal{L}_{\rho\gamma} = - \frac{e}{2g_{\rho}} F^{\mu\nu}
  \rho^0_{\mu\nu},
\end{equation}
where $F^{\mu\nu} = \partial_{\mu}A_{\nu} - \partial_{\nu}A_{\mu}$ is
the electromagnetic field strength tensor and $\rho^0_{\mu\nu} =
\partial_{\mu}\rho^0_{\nu} - \partial_{\nu}\rho^0_{\mu}$.  From the
width of the $\rho\rightarrow e^+ e^-$ decay, the value $g_{\rho} =
4.96$ is obtained.

\begin{figure}[htb]
  \begin{center}
  \includegraphics[width=13cm]{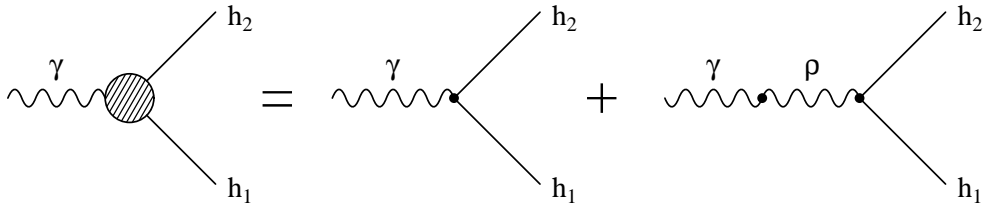}
  \caption{\label{fig:hhgamma} According to the vector meson dominance
    (VMD) model applied in this paper, the full electromagnetic vertex
    is a sum of the direct photon term and the $\rho$ meson
    contribution.}
  \end{center}
\end{figure}

In addition we have to specify the coupling of various hadrons to the
$\rho^0$. Hadrons can also directly couple to the electromagnetic field
$A^{\mu}$. The full electromagnetic vertex of hadrons $h_1$ and $h_2$
is, therefore, the sum of the direct photon term and the VMD
contribution (see Fig.~\ref{fig:hhgamma}). The vertex function
corresponding to the VMD contribution to the $h_1h_2\gamma$ coupling has
the form
\begin{equation}
  \label{eq:VMDvertex}
  V^{\mu\ldots}_{h_1h_2\gamma,\text{VMD}}(k) = F_{\text{VMD}}(k^2)
  V^{\mu\ldots}_{h_1h_2\rho}(k),
\end{equation}
where the VMD form factor appearing on the right hand side is given by
\begin{equation}
  \label{eq:VMDformfac}
  F_{\text{VMD}}(k^2) = -\frac{e}{g_{\rho}}
  \frac{k^2}{k^2-m_{\rho}^2+i\sqrt{k^2}\Gamma_{\rho}(k^2)},
\end{equation}
and is the product of the $\rho$
meson propagator and the $\rho\gamma$ vertex contribution. In
Eq.~(\ref{eq:VMDvertex}) $k$ is the photon four-momentum, $\mu$ is the
Lorentz index of the photon line and the dots stand for possible further
Lorentz indices corresponding to Rarita-Schwinger fields in case $h_1$
or $h_2$ are higher spin baryons.

For the electromagnetic interaction of a baryon resonance $R$
($h_1=R$ and $h_2=N$), the $g_{RN\rho}$ coupling constants can be
determined from the $R\rightarrow N\rho$ width of the baryon resonance
$R$. The VMD form factor in Eq.~(\ref{eq:VMDformfac}) is proportional to
$k^2$, therefore the VMD part of the electromagnetic vertex does not
contribute to the $R\rightarrow N\gamma$ decay width for real photons,
$k^2=0$. Thus, $g_{RN\gamma}$ can be fixed independently using the
photonic decay width $\Gamma_{R\rightarrow N\gamma}$. This is an
advantage of the choice of the VMD Lagrangian Eq.~(\ref{eq:VMD}). If
instead one uses the more common form
\begin{equation}
  \tilde{\mathcal{L}}_{\rho\gamma} = - \frac{em_{\rho}^2}{g_{\rho}}
  \rho^0_{\mu}A^{\mu},
\end{equation}
$k^2$ in the numerator of the VMD form factor, Eq.~(\ref{eq:VMDformfac})
is replaced by $m_{\rho}^2$. In that case the VMD contribution to
$\Gamma_{R\rightarrow N\gamma}$ is nonzero, and in fact overpredicts the
physical $N\gamma$ width for most of the baryon resonances, as pointed
out in Ref.~\cite{Friman_Pirner}.

\subsection{\label{sec:Nonresonant}Nonresonant contributions} 
\subsubsection{Contributions of direct photon couplings} 
In order to calculate the nonresonant Feynman diagrams
Fig.~\ref{fig:diagrams}(a)--(e), we have to specify the hadronic
and electromagnetic interaction Lagrangians of pions and nucleons.
We use a pseudovector $NN\pi$ coupling,
\begin{equation}
  \label{eq:L_NNpi}
  \mathcal{L}_{NN\pi} = - \frac{f_{NN\pi}}{m_{\pi}} \bar{\psi}_N\gamma_5
  \gamma^{\mu}\vec{\tau}\psi_N \cdot \partial_{\mu}\vec{\pi}.
\end{equation}
Following Ref.~\cite{Garcilazo} we use the value $f_{NN\pi} = 0.97$ for
the coupling constant.

The electromagnetic interaction Lagrangians must be chosen in such a way
that electromagnetic gauge invariance is fulfilled. This will ensure
that the photon field $A^{\mu}$ will couple to conserved currents
constructed from the hadron fields, and the resulting hadronic matrix
elements will satisfy the condition
$\mathcal{M}_{\mu}^{\text{had}}k^{\mu} = 0$. An important consequence
is, that the photon propagator can be written as $-i g_{\mu\nu}/k^2$,
which has been used in the derivation of Eq.~\ref{eq:matrixelement}.

Gauge invariant Lagrangians can be obtained by replacing derivatives
$\partial_{\mu}$ with the covariant derivative
\begin{equation}
  \nabla_{\mu} = \partial_{\mu} + i e A_{\mu} Q
\end{equation}
in all terms of the Lagrangian ($Q$ is the electric charge
operator). Carrying out this replacement in the nucleon kinetic energy
term results in the $NN\gamma$ interaction Lagrangian
$-e\bar{\psi}_N\slashed{A}Q\psi_N$. This is supplemented by the magnetic
term, which contains the field tensor $F^{\mu\nu}$, and is gauge
invariant. The complete $NN\gamma$ interaction is then
\begin{equation}
  \label{eq:L_NNgamma}
  \mathcal{L}_{NN\gamma} = 
  - e\bar{\psi}_N\left[\frac{1+\tau_3}{2}\slashed{A} -
    \left(\frac{1+\tau_3}{2}\kappa_p+\frac{1-\tau_3}{2}\kappa_n\right)
    \frac{\sigma_{\mu\nu}}{4m_N}F^{\mu\nu}\right]\psi_N.
\end{equation}
(The isospin 1/2 representation of the electric charge operator, $Q =
(1+\tau_3)/2$ has been substituted.)

Starting from the pion kinetic energy term we obtain the $\pi\pi\gamma$
interaction in the form
\begin{equation}
  \label{eq:L_gammapipi}
  \mathcal{L}_{\pi\pi\gamma} = -eA_{\mu}J_{\pi}^{\mu},
\end{equation}
where $J_{\pi}^{\mu} = i(\pi^{-}\partial^{\mu}\pi^{+} -
\pi^{+}\partial^{\mu}\pi^{-})$ is the pion current. In addition a
$\pi\pi\gamma\gamma$ term is also generated, but it does not contribute
to the studied process.

Inserting the covariant derivative in the pseudovector pion-nucleon
coupling term we obtain an $NN\pi\gamma$ contact interaction of the form
\begin{equation}
  \label{eq:L_NNpigamma}
  \mathcal{L}_{NN\pi\gamma} = - \frac{i e f_{NN\pi}}{m_{\pi}}
  \bar{\psi}_N\gamma_5 \gamma^{\mu}\vec{\tau}\psi \cdot
  A_{\mu}Q\vec{\pi}.
\end{equation}

Using the Lagrangians
Eqs.~(\ref{eq:L_NNpi}),(\ref{eq:L_NNgamma})--(\ref{eq:L_NNpigamma}) we
can calculate those Born contributions [diagrams (a)--(d) in
Fig.~\ref{fig:diagrams}] to $\mathcal{M}_{\mu}^{\text{had}}$ that
contain a direct photon coupling. The construction of the Lagrangians
assures that the sum of these contributions satisfies the gauge
invariance condition $\mathcal{M}_{\mu}^{\text{had}}k^{\mu}=0$. Note, however,
that the individual Feynman diagrams are not gauge invariant.

In order to describe the off-shell behavior of internal hadron lines we
apply at all hadronic vertices form factors given by
\begin{eqnarray}
  F_1(s) & = & \frac{1}{1 + (s-m_N^2)^2/\Lambda^4}, \label{eq:F1} \\
  F_2(u) & = & \frac{1}{1 + (u-m_N^2)^2/\Lambda^4}, \label{eq:F2} \\
  F_3(t) & = & \frac{1}{1 + (t-m_{\pi}^2)^2/\Lambda^4} \label{eq:F3}
\end{eqnarray}
for $s$-, $u$- and $t$-channel diagrams, respectively. These satisfy
\begin{equation}
  F_1(m_N^2) = F_2(m_N^2) = F_3(m_{\pi}^2) = 1.
\end{equation}

The application of different form factors to the individual diagrams
Fig.~\ref{fig:diagrams}(a)--(c) destroys the overall gauge invariance of the
Born contributions. The solution to this problem has been given by
Davidson and Workman in the case of pion photoproduction
\cite{Davidson-Workman}, and the method can be generalized to the
present case. We first write $\mathcal{M}_{\mu}^{\text{had}}$ in the
form
\begin{equation}
  \mathcal{M}_{\mu}^{\text{had}} = \bar{u}_f T_{\mu} u_i.
\end{equation}
Let $T_{\mu}^{\text{Born}}$ denote the Born contribution to $T_{\mu}$
obtained from direct photon terms. It can be shown by explicit
calculation of $T_{\mu}^{\text{Born}}$ from the Born channel Feynman
diagrams, that the replacement $T_{\mu}^{\text{Born}} \rightarrow
T_{\mu}^{\text{Born}} + \Delta T_{\mu}^{\text{Born}}$ makes the hadronic
matrix element $\mathcal{M}_{\mu}^{\text{had}}$ gauge invariant, if 
\begin{equation}
  \Delta T_{\mu}^{\text{Born}} = 
\frac{\sqrt{2}ef_{NN\pi}}{m_{\pi}} 2m_N
  \gamma_5\left[
    \left(\hat{F}(s,u,t)-F_3(t)\right) \frac{2q^{\mu}-k^{\mu}}{t - m_{\pi}^2} -
    \left(\hat{F}(s,u,t)-F_2(u)\right) \frac{2p_i^{\mu}-k^{\mu}}{u - m_N^2}
    \right],
\end{equation}
where
\begin{equation}
\label{eq:Fhat}
  \hat{F}(s,u,t) = F_1(s) + F_2(u) + F_3(t)
  - F_1(s)F_2(u) - F_1(s)F_3(t) - F_2(u)F_3(t)
  + F_1(s)F_2(u)F_3(t).
\end{equation}
$\hat{F}(s,u,t)$ was chosen in such a way that 
\begin{equation}
  \hat{F}(m_N^2,u,t) = \hat{F}(s,m_N^2,t) = \hat{F}(s,u,m_{\pi}^2) = 1,
\end{equation}
which means that the poles of $T_{\mu}^{\text{Born}}$ at $t=m_{\pi}^2$ and $u=m_N^2$
are canceled by the factors $\hat{F} - F_{2(3)}$. This means that the
term $\Delta T_{\mu}^{\text{Born}}$ can be generated by adding a
suitably chosen contact interaction to the Lagrangian.

Gauge invariance of the resulting $T_{\mu}^{\text{Born}}$ can be made
transparent by writing it in the form
\begin{equation}
  T_{\mu}^{\text{Born}} = \sum_{i=1}^4 A_i M_{i,\mu},
\end{equation}
where $M_{i,\mu}$ denote the gauge invariant combinations
\begin{eqnarray}
  M_{1,\mu} & = & \gamma_5\left(\gamma_{\mu}\slashed{k}-k_{\mu}\right), \label{eq:M1} \\
  M_{2,\mu} & = & \frac{\gamma_5}{2}\left[(2p_{i\mu}-k_{\mu})(2q\cdot k-M^2)
    - (2q_{\mu}-k_{\mu})(2p_i\cdot k-M^2)\right], \label{eq:M2} \\
  M_{3,\mu} & = & \frac{\gamma_5}{2}\left[\gamma_{\mu}(2p_f\cdot k+M^2)
    - (2p_{f\mu}+k_{\mu})\slashed{k}\right], \label{eq:M3} \\
  M_{4,\mu} & = & \frac{\gamma_5}{2}\left[\gamma_{\mu}(2p_i\cdot k-M^2)
    - (2p_{i\mu}-k_{\mu})\slashed{k}\right]. \label{eq:M4}
\end{eqnarray}
In the $k^2=0$ limit $M_{i,\mu}$ correspond to the gauge invariant
combinations defined in Ref.~\cite{Davidson-Workman} for the case of
pion photoproduction.


The coefficients $A_i$ are obtained from the explicit Feynman diagram
calculations and are given by
\begin{eqnarray}
  A_1 & = & - \frac{\sqrt{2}e f_{NN\pi}}{m_{\pi}} \left[
    \frac{1}{2m_N}\left(F_1\kappa_{n} + F_2\kappa_{p}\right)
    + \frac{2 m_N F_2}{u-m_N^2}(1+\kappa_{p})
    + \frac{2 m_N F_1}{s+m_N^2}\kappa_{n}
    \right], \label{eq:A1} \\
  A_2 & = & \frac{\sqrt{2}e f_{NN\pi}}{m_{\pi}}
  \frac{4 m_N \hat{F}}{(t-m_{\pi}^2)(u-m_N^2)}, \label{eq:A2} \\
  A_3 & = & \frac{\sqrt{2}e f_{NN\pi}}{m_{\pi}}
  \frac{2\kappa_{n}F_1}{s-m_N^2}, \label{eq:A3} \\
  A_4 & = & \frac{\sqrt{2}e f_{NN\pi}}{m_{\pi}}
  \frac{2\kappa_{p}F_2}{u-m_N^2}. \label{eq:A4}
\end{eqnarray}
In the derivation of Eqs.~(\ref{eq:A1})--(\ref{eq:A4}) we have used the
fact, that $k_{\mu}l^{\mu\nu}=0$, and thus arbitrary terms proportional
to $k_{\mu}$ can be added to $T_{\mu}$ without affecting the cross
section.

\subsubsection{VMD contributions to Born diagrams}
For the calculation of the VMD contributions we need the coupling of
hadrons to the $\rho^0$ meson. Here we face the same problems related to
gauge invariance as in the case of the direct photon couplings. First we
have to ensure that the relation $\mathcal{M}_{\mu}^{\text{had}}k^{\mu}
= 0$ holds without the inclusion of hadronic form factors. One
possibility to fulfill this condition is to define the interaction of
$\rho$ mesons with other hadrons by replacing derivatives
$\partial_{\mu}$ in the hadronic Lagrangians with
\begin{equation}
  \label{eq:covar_deriv_rho}
  \nabla_{\mu} = \partial_{\mu} - i \tilde{g}_{\rho} \vec{\rho}_{\mu}\cdot\vec{T},
\end{equation}
where $\vec{T}$ denotes the generators of the isospin SU(2) group. This
method is inspired by an SU(2) gauge theory with $\rho$ mesons as gauge
bosons.

In this way an $NN\rho$ interaction term can be obtained from the
nucleon kinetic energy term. Similarly to the direct photon coupling, a
magnetic type term can be added to it, yielding the total $NN\rho$
interaction Lagrangian
\begin{equation}
  \label{eq:L_NNrho}
  \mathcal{L}_{NN\rho} = \frac{\tilde{g}_{\rho}}{2}\bar\psi_N
  \left(\vec{\slashed{\rho}} -
  \kappa_{\rho}\frac{\sigma_{\mu\nu}}{4m_N}\vec\rho^{\mu\nu}\right) \cdot \vec{\tau}
  \psi_N.
\end{equation}
The $\rho\pi\pi$ term is obtained from the pion kinetic energy term and
has the form
\begin{equation}
  \label{eq:L_rhopipi}
  \mathcal{L}_{\rho\pi\pi} =
  - \tilde{g}_{\rho}\left[(\partial^{\mu}\vec{\pi}) \times
    \vec{\pi}\right]\cdot\vec{\rho}_{\mu}.
\end{equation}
Comparing the Lagrangians Eqs.~(\ref{eq:L_NNrho}) and
(\ref{eq:L_rhopipi}) with the traditional forms of the $NN\rho$ and
$\rho\pi\pi$ couplings, we see that their construction in terms of the
covariant derivative Eq.~(\ref{eq:covar_deriv_rho}) provides a relation
of their coupling constants in the form
\begin{equation}
2g_{NN\rho} = g_{\rho\pi\pi} = \tilde{g}_{\rho}.
\end{equation}
From the width of the decay $\rho\rightarrow\pi\pi$ the value
$g_{\rho\pi\pi} = 5.96$ is obtained. $g_{NN\rho}$ can be determined from
low energy nucleon-nucleon scattering. In Ref.~\cite{Fernandez} the
value $g_{NN\rho} = 2.6$ was used, yielding the ratio
$g_{\rho\pi\pi}/g_{NN\rho} = 2.29$, which is reasonably close to the
value of 2 predicted by SU(2) gauge invariance. In the present
calculation we use the values $\tilde{g}_{\rho} = g_{\rho\pi\pi} = 5.96$
and $g_{NN\rho} = \tilde{g}_{\rho}/2 = 2.98$.

Inserting the covariant derivative Eq.~(\ref{eq:covar_deriv_rho}) in the
pseudovector $NN\pi$ Lagrangian Eq.~(\ref{eq:L_NNpi}) we obtain an
$NN\pi\rho$ contact interaction,
\begin{equation}
  \mathcal{L}_{NN\pi\rho} = - \frac{\tilde{g}_{\rho}f_{NN\pi}}{m_{\pi}}
  \bar\psi_N \gamma_5 \gamma^{\mu}\vec{\tau}\psi\cdot
  \left(\vec{\rho}_{\mu}\times\vec{\pi}\right).
\end{equation}

In accordance with Eq.~(\ref{eq:VMDvertex}) the VMD contribution to the
hadronic matrix element can be written in the form
\begin{equation}
  \mathcal{M}_{\mu}^{\text{had,VMD}} =
  F_{\text{VMD}}(k^2)\tilde{\mathcal{M}}_{\mu},
\end{equation}
where the VMD form factor $F_{\text{VMD}}(k^2)$ is given by
Eq.~(\ref{eq:VMDformfac}). Feynman diagrams representing
$\tilde{\mathcal{M}}_{\mu}$ can be obtained from the VMD diagrams by
truncating the dilepton part, starting from the $\rho$ propagator.

At hadronic vertices we employ the same form factors
[Eqs.~(\ref{eq:F1})--(\ref{eq:F3})] as in the direct photon
contributions.  Then we write $\tilde{\mathcal{M}}_{\mu}$ in the form
\begin{equation}
  \tilde{\mathcal{M}}_{\mu} = \bar{u}_f \tilde{T}_{\mu} u_i.
\end{equation}
The explicit form of $\tilde{T}_{\mu}^{\text{Born,VMD}}$ (the
contribution to $\tilde{T}_{\mu}$ of Born diagrams with VMD coupling) is
calculated from the relevant Feynman diagrams. We observe that the
replacement $\tilde{T}_{\mu}^{\text{Born,VMD}} \rightarrow
\tilde{T}_{\mu}^{\text{Born,VMD}} + \Delta
\tilde{T}_{\mu}^{\text{Born,VMD}}$ ensures the validity of the gauge
invariance relation, $\mathcal{M}_{\mu}^{\text{had,VMD}}k^{\mu} = 0$ if
$\Delta \tilde{T}_{\mu}^{\text{Born,VMD}}$ is chosen as
\begin{eqnarray}
  \lefteqn{\Delta\tilde{T}_{\mu}^{\text{Born,VMD}} =  
    \frac{\tilde{g}_{\rho}f_{NN\pi}}{\sqrt{2}m_{\pi}} 2m_N \gamma_5 } \\
  & & \times \left[
    \left(\hat{F}(s,u,t)-F_2(u)\right) \frac{2p_i^{\mu}-k^{\mu}}{u - m_N^2} -
    \left(\hat{F}(s,u,t)-F_1(u)\right) \frac{2p_f^{\mu}+k^{\mu}}{s - m_N^2} -
    2\left(\hat{F}(s,u,t)-F_3(t)\right) \frac{2q^{\mu}-k^{\mu}}{t - m_{\pi}^2}
    \right].
\end{eqnarray}
This $\Delta \tilde{T}_{\mu}^{\text{Born,VMD}}$ is free from poles, and
is assumed to be generated by suitable contact terms added to the
Lagrangian.

The obtained $\tilde{T}_{\mu}^{\text{Born,VMD}}$ can be expanded as
\begin{equation}
  \tilde{T}_{\mu}^{\text{Born,VMD}} = \sum_{i=1}^5 \tilde{A}_i M_{i,\mu},
\end{equation}
where $M_{1...4,\mu}$ are given in Eqs.~(\ref{eq:M1})--(\ref{eq:M4}),
and
\begin{equation}
 \label{eq:M5}
 M_{5,\mu} = \frac{\gamma_5}{2}\left[(2p_{f\mu}+k_{\mu})(2q\cdot k-M^2)
   - (2q_{\mu}-k_{\mu})(2p_f\cdot k+M^2)\right].
\end{equation}
The coefficients $\tilde{A}_i$ are obtained as
\begin{eqnarray}
  \tilde{A}_1 & = & \frac{\tilde{g}_{\rho}f_{NN\pi}}{\sqrt{2}m_{\pi}}
  \left[
    \frac{\kappa_{\rho}}{2m_N}\left(F_2 - F_1\right)
    + 2 m_N(1+\kappa_{\rho})\left( \frac{F_2}{u-m_N^2} -
    \frac{F_1}{s-m_N^2} \right)
    \right], \\
  \tilde{A}_2 & = &
  - \frac{\tilde{g}_{\rho}f_{NN\pi}}{\sqrt{2}m_{\pi}} \frac{4 m_N
    \hat{F}}{(t-m_{\pi}^2)(u-m_N^2)}, \\
  \tilde{A}_3 & = &
  \frac{\tilde{g}_{\rho}f_{NN\pi}}{\sqrt{2}m_{\pi}}
  \frac{2\kappa_{\rho}F_1}{s-m_N^2}, \\
  \tilde{A}_4 & = &
  - \frac{\tilde{g}_{\rho}f_{NN\pi}}{\sqrt{2}m_{\pi}}
  \frac{2\kappa_{\rho}F_2}{u-m_N^2}, \\
  \tilde{A}_5 & = &
  - \frac{\tilde{g}_{\rho}f_{NN\pi}}{\sqrt{2}m_{\pi}} \frac{4 m_N
    \hat{F}}{(t-m_{\pi}^2)(s-m_N^2)}.
\end{eqnarray}

\subsubsection{$t$-channel $\rho$- and $a_1$-exchange contributions}

We also calculated the contributions of the $t$-channel $\rho$- and
$a_1$-exchange diagrams, Fig.~\ref{fig:diagrams}(e). For the
$\rho$ exchange we adopt the $\rho\pi\gamma$ interaction Lagrangian from
Ref.~\cite{Feuster_Mosel_NPA},
\begin{equation}
  \mathcal{L}_{\rho\pi\gamma} = e\frac{g_{\rho\pi\gamma}}{4m_{\pi}}
  \epsilon_{\mu\nu\lambda\sigma} F^{\mu\nu}\vec{\rho}^{\lambda\sigma}
  \cdot \vec{\pi}.
\end{equation}
The value of the coupling constant, $g_{\rho\pi\gamma} = 0.103$, is
obtained from the width of the $\rho \rightarrow \pi\gamma$ decay.
Lagrangians equivalent to the above $\mathcal{L}_{\rho\pi\gamma}$ have
been used in Refs.~\cite{Fernandez,Garcilazo}.

The $a_1\pi\gamma$ interaction was studied in Ref.~\cite{Xiong}. In that
paper the momentum space form of the interaction Lagrangian was
given. Its coordinate space equivalent is given by
\begin{equation}
  \label{eq:a1pigamma}
  \mathcal{L}_{a_1\pi\gamma} = - i e\frac{g_{a_1\pi\gamma}}{m_{\pi}}
  \vec{a}_{\mu} F^{\mu\nu} \cdot \partial_{\nu}\vec{\pi},
\end{equation}
where $\vec{a}_{\mu}$ denotes the axial-vector--isovector $a_1$
field. From the width of the $a_1 \rightarrow \pi\gamma$ decay we get
$g_{a_1\pi\gamma} = 0.106$ for the coupling constant.

We also need to specify the form of the $NNa_1$ interaction. The role of
$t$-channel $a_1$ exchange in the nucleon-nucleon interaction was
studied in Ref.~\cite{Durso}. They take the $NNa_1$ Lagrangian from the
chiral $SU(2) \times SU(2)$ model of Ref.~\cite{Wess-Zumino}. In that
model the Lagrangian has the form
\begin{equation}
  \label{eq:L_NNa1}
  \mathcal{L}_{NNa_1} = g_{NNa_1}\bar\psi_N
  \gamma^{\mu}\gamma_5\vec{\tau}\psi_N \cdot \vec{a_{\mu}},
\end{equation}
and the coupling constant is related to the pseudovector pion-nucleon
coupling via
\begin{equation}
  \frac{g_{NNa_1}}{m_{a_1}} = \frac{f_{NN\pi}}{m_{\pi}}.
\end{equation}
This relation gives the value $g_{NNa_1} = 8.65$. In Ref.~\cite{Yu} the
nucleon-$a_1$ coupling has been determined from the nucleon axial form
factor, and the value $g_{NNa_1} = 6.7$ was obtained.

In close analogy with the Born contributions we apply form factors given by
\begin{equation}
  F_V(t) = \frac{1}{1 + (t-m_V^2)^2/\Lambda^4}
\end{equation}
for $t$-channel $\rho$- and $a_1$-exchange diagrams, where $m_V$ denotes
the $\rho$ or $a_1$ meson mass.

We found that the contribution of $t$-channel $\rho$ exchange is at
least three orders of magnitude smaller than the Born contribution in
the $\sqrt{s}\le1$GeV energy range. The $a_1$-exchange contribution is
even smaller and never exceeds 10\% of the $\rho$ contribution.

\subsection{Contributions of baryon resonances} 
\subsubsection{Interaction Lagrangians}

In order to calculate the $s$- and $u$-channel baryon resonance
contributions, diagrams Fig.~\ref{fig:diagrams}(f) and (g), we have
to specify the coupling of baryon resonances to the $\pi N$, $\rho N$
and $\gamma N$ channels.

Similarly to the nucleon-pion interaction we employ pseudovector
couplings in the case of spin-1/2 nucleon resonances,
\begin{equation}
  \mathcal{L}_{R_{1/2}N\pi} = - \frac{g_{RN\pi}}{m_{\pi}} \bar{\psi}_R \Gamma
  \gamma^{\mu}\vec{\tau}\psi_N \cdot \partial_{\mu}\vec{\pi}
  + \text{H.c.} \label{eq:R12Npi}
\end{equation}
In the spin-3/2 case we use the Lagrangian
\begin{equation}
  \mathcal{L}_{R_{3/2}N\pi} = \frac{g_{RN\pi}}{m_{\pi}}
  \bar{\psi}_R^{\mu} \Gamma \vec{\tau}\psi_N\cdot\partial_{\mu}\vec{\pi}
  + \text{H.c.}, \label{eq:R32Npi}
\end{equation}
while in the spin-5/2 case the Lagrangian
\begin{equation}
  \mathcal{L}_{R_{5/2}N\pi} = \frac{g_{RN\pi}}{m_{\pi}}
  \bar{\psi}_R^{\mu\nu} \Gamma \vec{\tau}\psi_N \cdot
  \partial_{\mu}\partial_{\nu}\vec{\pi} + \text{H.c.} \label{eq:R52Npi}
\end{equation}
In the above $\Gamma = \gamma_5$ for $J^P = {\frac{1}{2}}^{+}$,
${\frac{3}{2}}^{-}$ and ${\frac{5}{2}}^{+}$ resonances and $\Gamma = 1$
otherwise. $\psi_{R}^{\mu}$ and $\psi_{R}^{\mu\rho}$ are the
Rarita-Schwinger fields describing spin-$\frac{3}{2}$ and $\frac{5}{2}$
resonances, respectively, and $\vec{\tau}$ are the (isospin) Pauli
matrices. In the case of $\Delta$ resonances $\vec{\tau}$ has to be
replaced by the isospin $\frac{3}{2} \rightarrow \frac{1}{2}$ transition
matrices, $\vec{T}$.

We now list the Lagrangians describing the $RN\gamma$ and $RN\rho$
coupling of baryon resonances. For spin-$1/2$ nucleon resonances these
are given by
\begin{eqnarray}
  \mathcal{L}_{R_{1/2}N\gamma} & = & \frac{g_{RN\gamma}}{2m_{\rho}} \bar{\psi}_{R}
  \sigma^{\mu\nu} \tilde{\Gamma} \psi_N F_{\mu\nu} +
  \text{H.c.}, \label{eq:RNgamma_first}\\
  \mathcal{L}_{R_{1/2}N\rho} & = & \frac{g_{RN\rho}}{2m_{\rho}} \bar{\psi}_{R}
  \vec{\tau} \sigma^{\mu\nu} \tilde{\Gamma} \psi_N \cdot \vec{\rho}_{\mu\nu} +
  \text{H.c.}
\end{eqnarray}
For spin-$3/2$ nucleon resonances the corresponding Lagrangians are
\begin{eqnarray}
  \mathcal{L}_{R_{3/2}N\gamma} & = & - \frac{ig_{RN\gamma}}{m_{\rho}}
  \bar{\psi}_{R}^{\mu} \gamma^{\nu}\tilde{\Gamma} \psi_N
  F_{\mu\nu} + \text{H.c.}, \\
  \mathcal{L}_{R_{3/2}N\rho} & = & - \frac{ig_{RN\rho}}{m_{\rho}}
  \bar{\psi}_{R}^{\mu}\vec{\tau} \gamma^{\nu}\tilde{\Gamma} \psi_N \cdot
  \vec{\rho}_{\mu\nu} + \text{H.c.},
\end{eqnarray}
and for spin-$5/2$ nucleon resonances we use
\begin{eqnarray}
  \mathcal{L}_{R_{5/2}N\gamma} & = & - \frac{ig_{RN\gamma}}{m_{\rho}}
  \bar{\psi}_{R}^{\mu\rho} \gamma^{\nu}\tilde{\Gamma} (\partial_{\rho}\psi_N)
  F_{\mu\nu} + \text{H.c.}, \\
  \mathcal{L}_{R_{5/2}N\rho} & = & - \frac{ig_{RN\rho}}{m_{\rho}}
  \bar{\psi}_{R}^{\mu\rho}\vec{\tau} \gamma^{\nu}\tilde{\Gamma}
  (\partial_{\rho}\psi_N) \cdot
  \vec{\rho}_{\mu\nu} + \text{H.c.} \label{eq:RNgamma_last}
\end{eqnarray}
For the $RN\rho$ couplings, $\vec{\tau}$ is replaced by $\vec{T}$ in the
case of $\Delta$ resonances, similarly to the $RN\pi$ case. In
Eqs.~(\ref{eq:RNgamma_first})--(\ref{eq:RNgamma_last}) $\tilde{\Gamma} =
\gamma_5$ for $J^P = {\frac{1}{2}}^{-}$, ${\frac{3}{2}}^{+}$ and
${\frac{5}{2}}^{-}$ resonances and $\tilde{\Gamma} = 1$ otherwise.

Dilepton production in the Dalitz decay of baryon resonances
($R\rightarrow Ne^+e^-$) was studied in Refs.~\cite{Krivoruchenko} and
\cite{Zetenyi}. In \cite{Zetenyi} we discussed the possible forms of
matrix elements of the electromagnetic current between a resonance and a
nucleon state. We demonstrated that the contributions of the various
possibilities do not differ significantly, unless the resonance mass is
far from the nominal value. Based on this result, the matrix elements
containing the lowest power of external momenta were chosen for the
calculation of the resulting dilepton spectra.  The Lagrangians
Eqs.~(\ref{eq:RNgamma_first})--(\ref{eq:RNgamma_last}) correspond to the
same choice in the sense that the matrix elements calculated from them
coincide with those chosen in Ref.~\cite{Zetenyi}.

\subsubsection{\label{sec:propa_FF} Propagators and form factors}
The propagator of spin-3/2 baryon resonances is
\begin{equation}
G_{R_{3/2}}^{\mu\nu}(p) = \frac{i}{p^2-m_R^2+i\sqrt{p^2}\Gamma_R(p^2)}
P_{3/2}^{\mu\nu}(p,m_R),
\end{equation}
where 
\begin{equation}
  P_{3/2}^{\mu\nu}(p,m_R) = -(\slashed{p}+m_R)
  \left(g^{\mu\nu} -
  \frac{\gamma^{\mu}\gamma^{\nu}}{3} -
  \frac{2}{3}\frac{p^{\mu}p^{\nu}}{m_R^2} + 
  \frac{p^{\mu}\gamma^{\nu} - p^{\nu}\gamma^{\mu}}{3m_R}
  \right).
\end{equation}
On the mass-shell $P_{3/2}^{\mu\nu}(p,m_R)$ coincides with
the spin-3/2 projector operator. 

For the spin-5/2 propagator we use
\begin{equation}
G_{R_{5/2}}^{\mu\nu,\rho\sigma}(p) =
\frac{i}{p^2-m_R^2+i\sqrt{p^2}\Gamma_R(p^2)}
P_{5/2}^{\mu\nu,\rho\sigma}(p,m_R),
\end{equation}
where 
\begin{equation}
P_{5/2}^{\mu\nu,\rho\sigma}(p,m_R) = (\slashed{p} + m_R)
\left[ \frac{3}{10}\left(G^{\mu\rho}G^{\nu\sigma}
                       + G^{\mu\sigma}G^{\nu\rho}
                   \right)
     - \frac{1}{5}G^{\mu\nu}G^{\rho\sigma}
     - \frac{1}{10}\left(T^{\mu\rho}G^{\nu\sigma}
                       + T^{\nu\sigma}G^{\mu\rho}
                       + T^{\mu\sigma}G^{\nu\rho}
                       + T^{\nu\rho}G^{\mu\sigma}
                   \right)
\right],
\end{equation}
with
\begin{equation}
G^{\mu\nu} = - g^{\mu\nu} + \frac{p^{\mu}p^{\nu}}{m_R^2},
\end{equation}
and
\begin{equation}
T^{\mu\nu} =
- \frac{1}{2}(\gamma^{\mu}\gamma^{\nu}-\gamma^{\nu}\gamma^{\mu})
+ \frac{p^{\mu}\left(\slashed{p}\gamma^{\nu}
                   - \gamma^{\nu}\slashed{p}
               \right)}{2m_R^2}
- \frac{p^{\nu}\left(\slashed{p}\gamma^{\mu}
                   - \gamma^{\mu}\slashed{p}
               \right)}{2m_R^2}.
\end{equation}

We parametrize the $p^2$ dependence of the $N\pi$ and $N\eta$ width of
baryon resonances as \cite{Teis97}
\begin{equation}
  \label{eq:Gamma_p}
  \Gamma(p^2) = \Gamma(m_R^2)\frac{m_R}{\sqrt{p^2}}
  \left(\frac{q}{q_R}\right)^{2l+1}
  \left(\frac{q_R^2+\delta^2}{q^2+\delta^2}\right)^{l+1},
\end{equation}
where $l$ is the angular momentum of the pion or $\eta$ meson, $q$ is
the magnitude of the outgoing three-momentum in the rest frame of the
decaying resonance given by
\begin{equation}
  q = \frac{\sqrt{\lambda(p^2,m_N^2,m_{\pi(\eta)}^2)}}{2\sqrt{p^2}},
\end{equation}
while $q_R$ is the same quantity for an on-shell resonance, $p^2 =
m_R^2$. The cutoff parameter $\delta$ is given by
\begin{equation}
  \delta^2 = \left(m_R - m_N - m_{\pi(\eta)} \right)^2 +
  \frac{\left[\Gamma(m_R^2)\right]^2}{4},
\end{equation}
with the exception of the $\Delta(1232)$ where the value $\delta =
0.3$GeV, and the $N(1535)$, where $\delta = 0.5$GeV has been used.

The $p^2$ dependence of the $N\pi$ width of baryon resonances can be
calculated from the appropriate Feynman diagrams using the effective
Lagrangians Eqs.~(\ref{eq:R12Npi})--(\ref{eq:R52Npi}). In order to get a
$p^2$ dependence numerically similar to Eq.~(\ref{eq:Gamma_p}) we employ
a cutoff factor of the form
\begin{equation}
  F(p^2) = \sqrt{\frac{m_R}{\sqrt{p^2}}}
  \left(\frac{q_R^2+\delta^2}{q^2+\delta^2}\right)^\frac{l+1}{2}
\end{equation}
at each $RN\pi$ vertex.

Two-pion decays of baryon resonances are assumed to proceed through an
intermediate baryon or meson resonance, as $R \rightarrow
(\Delta/N(1440))\pi \rightarrow N\pi\pi$ or $R \rightarrow N(\rho/\sigma)
\rightarrow N\pi\pi$. For the $p^2$ dependence of the corresponding
decay width we choose the expression obtained from a Feynman diagram
calculation, multiplied by the cutoff factor
\begin{equation}
  F_{\pi\pi}(p^2) = \left[\frac{(\sqrt{p^2}-m_N-2m_{\pi})^2 + \delta^2}
    {(m_R-m_N-2m_{\pi})^2 + \delta^2} \right]^2.
\end{equation}

It was pointed out in Refs.~\cite{Garcilazo,Feuster_Mosel_NPA} that the
pion photoproduction data can be reproduced only if the $u$-channel
resonance diagrams are multiplied by the extra cutoff factor
$\Lambda_u^2/(\Lambda_u^2 + q^2)$, with $\Lambda_u = $ 0.3~GeV. ($q$ is
the magnitude of the pion momentum in the center-of-mass frame.) The
role of this cutoff is to remove the high-energy divergence of these
contributions. A similar divergence of the $u$-channel contributions
occurs in the case of the $\pi N \rightarrow Ne^+e^-$ process discussed
in the present paper. However, we find that diagrams with higher spin
resonances diverge faster. Especially spin-5/2 resonances need a
stronger cutoff. Therefore we use a spin dependent cutoff factor of the
form
\begin{equation}
  \label{eq:uch_cutoff}
  F_u(p^2) = \left(\frac{\Lambda_u^2}{\Lambda_u^2 + q^2}\right)^{J}
\end{equation}
for $u$-channel diagrams with a resonance of spin $J$. Following
Refs.~\cite{Garcilazo,Feuster_Mosel_NPA} we use the value $\Lambda_u = $
0.3~GeV for the cutoff parameter.

\section{\label{sec:resparam} Resonance parameters}
Our model includes 16 baryon resonances below 2 GeV that have three- or
four-star status according to the Review of Particle Physics \cite{PDG}.
We did not include the state $\Delta(1920)$ because we have no
information about its $N\rho$ and $N\gamma$ branching ratio. We also
excluded the spin-7/2 $\Delta(1950)$ resonance.

We take the mass and total width of the resonances from \cite{PDG}. The
$RN\pi$ and $RN\rho$ coupling constants are determined from the partial
decay widths, that are obtained from the total width and the mean value
of the lower and upper bounds of the appropriate branching ratio listed
in \cite{PDG}. 

The $N\gamma$ branching ratios are poorly known for most
resonances. Also, the partial decay widths give no information about the
sign of coupling constants. These signs determine the signs of
interference terms in the $\pi N \rightarrow Ne^+e^-$ cross section. To
overcome these problems we also calculated the total cross section of
pion photoproduction, fitted to the available experimental data, using
the $RN\gamma$ coupling constants as fit parameters. We varied also the
signs of these coupling constants.

The Feynman diagrams contributing to pion photoproduction can be
obtained from the time inverse of the diagrams in
Fig.~\ref{fig:diagrams} by truncating at the photon propagator. The
calculation of the matrix elements goes along the same lines as for
the $\pi N \rightarrow Ne^+e^-$ process, but now the photon is on-shell,
$k^2 = 0$, which substantially simplifies the obtained expressions. In
particular, there are no VMD contributions to pion photoproduction
because of the choice of the $\rho\gamma$ Lagrangian of the form
Eq.~(\ref{eq:VMD}).

Nonresonant contributions are calculated according to the
gauge-invariance preserving scheme of Ref.~\cite{Davidson-Workman},
which can be obtained from the formulas of
Sec.~\ref{sec:Nonresonant} in the $k^2 = 0$ limit. Resonant
contributions are calculated numerically.

During the fitting procedure we varied the $RN\gamma$ coupling constants
within the ranges allowed by the total width and $N\gamma$ branching
ratios of the resonance $R$ as listed in \cite{PDG}. An exception is the
$N(1680)$ resonance where we reduced the limits of the $p\gamma$
branching ratio by a factor of about 10 to coincide with the limits of
the $n\gamma$ branching ratio. This reduction was necessary because
otherwise the large $N(1680)p\gamma$ coupling resulted in a high
$N(1680)$ peak on the $\gamma p \rightarrow \pi^{+}n$ total cross
section starting with a rapid rise already below 1~GeV laboratory photon
energy, which is not seen in experimental data. The other exception is
the $\Delta(1232)$ where we decreased the photonic branching ratio by
about 25\% below the PDG lower bound in order to obtain a reasonable
description of the pion photoproduction data.

We repeated the fit with various values of the cutoff parameter
$\Lambda$ of the Born contributions. The best fit was obtained with the
value $\Lambda = 0.63$ GeV.

\begin{figure}[htb]
  \begin{center}
  \includegraphics[width=5.9cm]{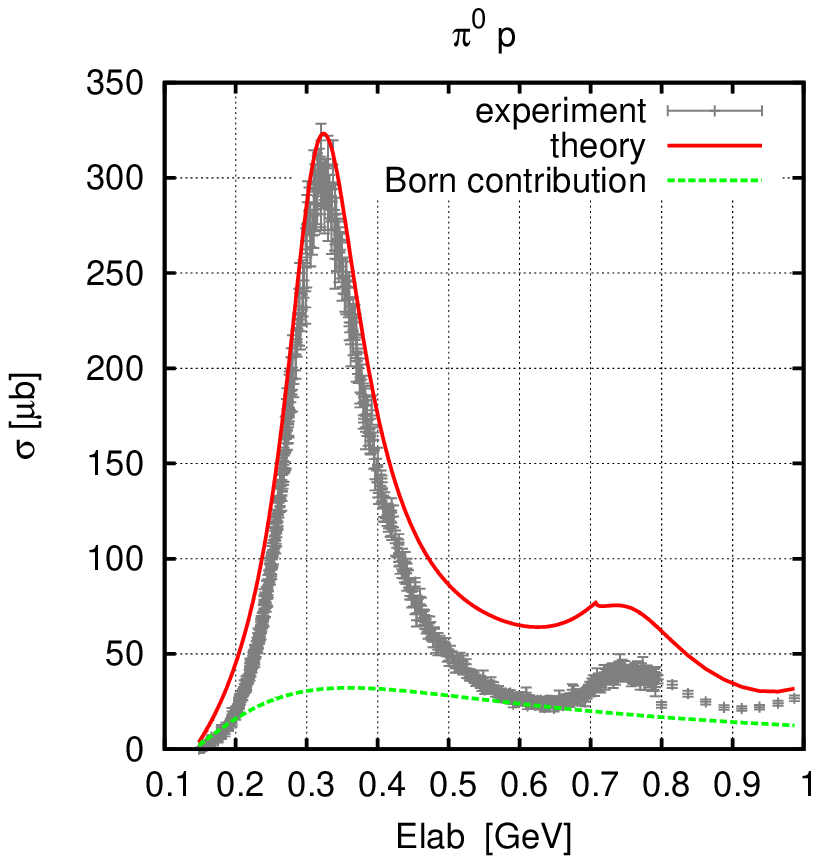}
  \includegraphics[width=5.9cm]{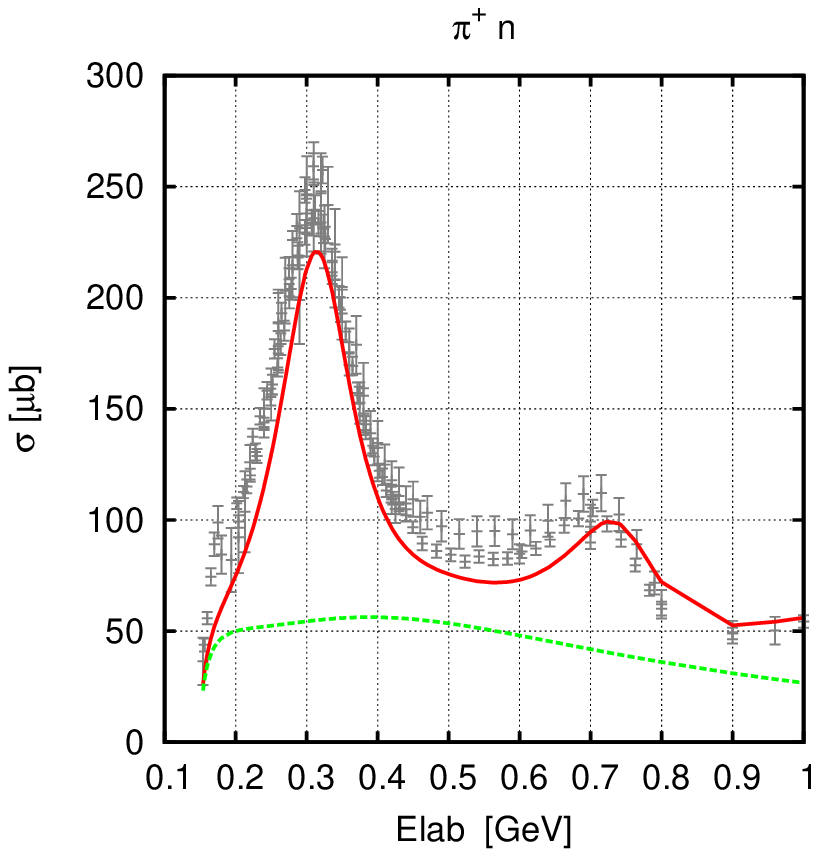}
  \includegraphics[width=5.9cm]{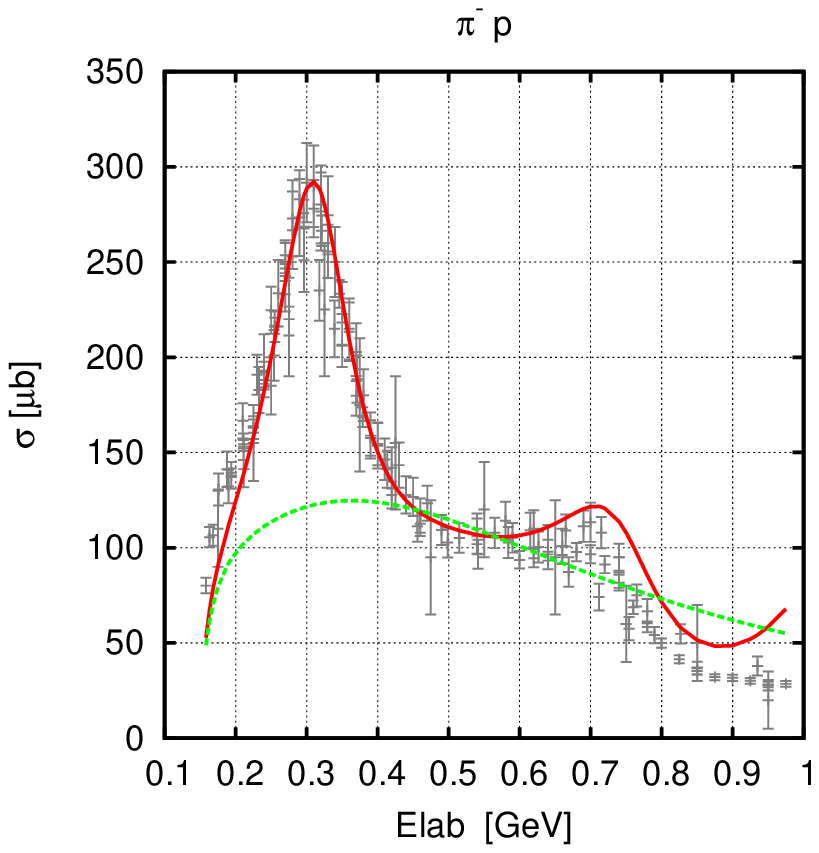}
  \caption{\label{fig:pion-photoprod} (Color online) Total cross section
    of pion photoproduction. Results of our calculation are compared
    with experimental data. Dashed lines show the contribution of Born
    diagrams.}
  \end{center}
\end{figure}

The resonance parameters---including the fitted $RN\gamma$ coupling
constants---are summarized in Table~\ref{tab:resonances}.
Figure~\ref{fig:pion-photoprod} shows the total pion photoproduction
cross sections calculated from our best fit in comparison with the
experimental data.  We also show the contribution of Born diagrams. The
three plots correspond to the processes $\gamma p\rightarrow\pi^0 p$,
$\gamma p\rightarrow\pi^+ n$, and $\gamma n\rightarrow\pi^- p$.

The discrepancies seen in the $\pi^0 p$ and $\pi^+ n$ channels are hard
to cure in the framework of the present model. Both cross sections
contain the $p\gamma$ coupling constant of each resonance in the
$s$-channel contributions. Thus the ratio of the contribution to the
$\pi^0 p$ and $\pi^+ n$ channels of each $s$-channel resonance diagram
is purely determined by isospin Clebsch-Gordan coefficients appearing in
the $RN\pi$ vertex. Since Born and $s$-channel resonance contributions
dominate the cross sections little freedom is left to balance the two
channels with $\gamma+p$ initial state.

In the $\pi^- p$ channel above 0.8~GeV laboratory photon energy the total
cross section is less than the Born contribution. This is a result of a
destructive interference.

\begin{table}
  \caption{\label{tab:resonances} Parameters of the 16 baryon resonances included in the model.}
  \begin{ruledtabular}
  \begin{tabular}{|c|c|d|d|r|r|d|d|dd|}
    & \raisebox{-2ex}[0pt][0pt]{$J^P$} & m_R & \Gamma_\text{tot} & \multicolumn{2}{c|}{BR (\%)} & \multicolumn{4}{c|}{coupling constants}\\ \cline{5-10}
    & & \text(GeV) & \text{(MeV)} & $N\pi$ & $N\rho$ & \multicolumn{1}{c|}{$g_{RN\pi}$} & \multicolumn{1}{c|}{$g_{RN\rho}$} & \multicolumn{1}{c}{$g_{Rp\gamma}$} & \multicolumn{1}{c|}{$g_{Rn\gamma}$} \\
    \hline
    $\Delta$(1232) & $3/2^+$ & 1.232 & 0.118 &  100 &  0  & 1.52   & 0     &  -1.5     &   -1.5    \\
    $N(1440)$      & $1/2^+$ & 1.440 & 0.3   &   65 &  0  & 7.40   & 0     &  0.204    &   -0.088  \\
    $N(1520)$      & $3/2^-$ & 1.520 & 0.115 &   60 & 20  & 1.94   & 9.92  &  -0.67    &   0.654   \\
    $N(1535)$      & $1/2^-$ & 1.535 & 0.15  &   45 &  2  & 0.838  & 1.73  &  0.204    &   0.033   \\
    $N(1650)$      & $1/2^-$ & 1.655 & 0.165 &   77 &  8  & 1.09   & 0.994 &  -0.186   &   -0.181  \\
    $N(1675)$      & $5/2^-$ & 1.675 & 0.15  &   40 &  1  & 0.122  & 6.74  &  0.124    &   -0.679  \\
    $N(1680)$      & $5/2^+$ & 1.685 & 0.13  &   67 &  9  & 0.509  & 6.03  &  -0.38    &   -0.381  \\
    $N(1700)$      & $3/2^-$ & 1.700 & 0.10  &   10 & 17  & 0.434  & 1.25  &  -0.135   &   -0.060  \\
    $N(1710)$      & $1/2^+$ & 1.710 & 0.10  &   15 & 15  & 1.28   & 1.68  &  0.0694   &   0.044   \\
    $N(1720)$      & $3/2^+$ & 1.720 & 0.2   &   15 & 77  & 0.208  & 9.37  &  -0.045   &   0.515   \\
    $\Delta(1600)$ & $3/2^+$ & 1.600 & 0.35  &   17 & 12  & 0.355  & 16.5  &  0.189    &   0.189   \\
    $\Delta(1620)$ & $1/2^-$ & 1.630 & 0.145 &   25 & 16  & 0.587  & 1.72  &  0.0272   &   0.0272  \\
    $\Delta(1700)$ & $3/2^-$ & 1.700 & 0.3   &   15 & 42  & 0.922  & 3.40  &  0.361    &   0.531   \\
    $\Delta(1905)$ & $5/2^+$ & 1.890 & 0.33  &   12 & 60  & 0.178  & 4.76  &  0.173    &   0.173   \\
    $\Delta(1910)$ & $1/2^+$ & 1.910 & 0.25  &   22 &  0  & 1.95   & 0     &  0.165    &   0.165   \\
    $\Delta(1930)$ & $5/2^-$ & 1.960 & 0.36  &   10 &  0  & 0.0491 & 0     &  0.0      &   0.0  
  \end{tabular}
  \end{ruledtabular}
\end{table}

\section{\label{sec:results} Results for dilepton production}
\begin{figure}[htb]
  \begin{center}
  \includegraphics[width=8.6cm]{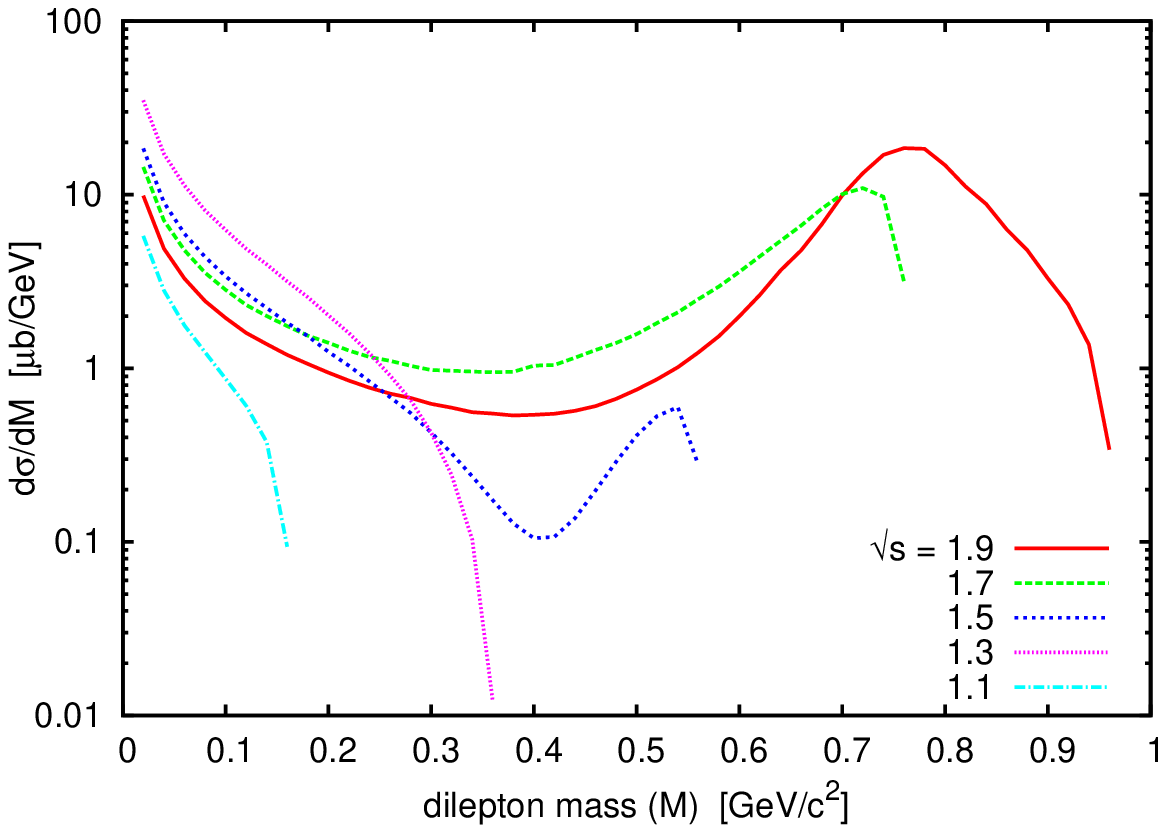}
  \caption{\label{fig:spectra} (Color online) Dilepton invariant mass
    spectra from the reaction $\pi^- + p \rightarrow n + e^+ + e^-$ for
    various collision energies.}
  \end{center}
\end{figure}

We used the effective field theory model described in
Sec.~\ref{sec:EFT} to calculate the matrix elements of the process
$\pi^- + p \rightarrow n + e^+ + e^-$ represented by the Feynman
diagrams of Fig.~\ref{fig:diagrams}. Then we used Eq.~(\ref{eq:dsdm}) to
calculate the differential cross section $d\sigma/dM$. The integrations
were carried out numerically using a Monte Carlo technique. The resulting
dilepton spectra for various collision energies are shown in
Fig.~\ref{fig:spectra}.  The mass spectra at 1.3 GeV and below are
monotonically decreasing, above 1.5 GeV pion energy the $\rho$ meson
contributes. At 1.5 and 1.7 GeV energy only the tail of the $\rho$ meson
spectrum is populated, still it produces a peak in the dilepton
invariant mass spectrum.  Note, however, that in the model no direct $\rho$
channel is included. The effect of the $\rho$ meson is encoded in the
VMD form factors of hadrons.

\begin{figure}[htb]
  \begin{center}
  \includegraphics[width=8.6cm]{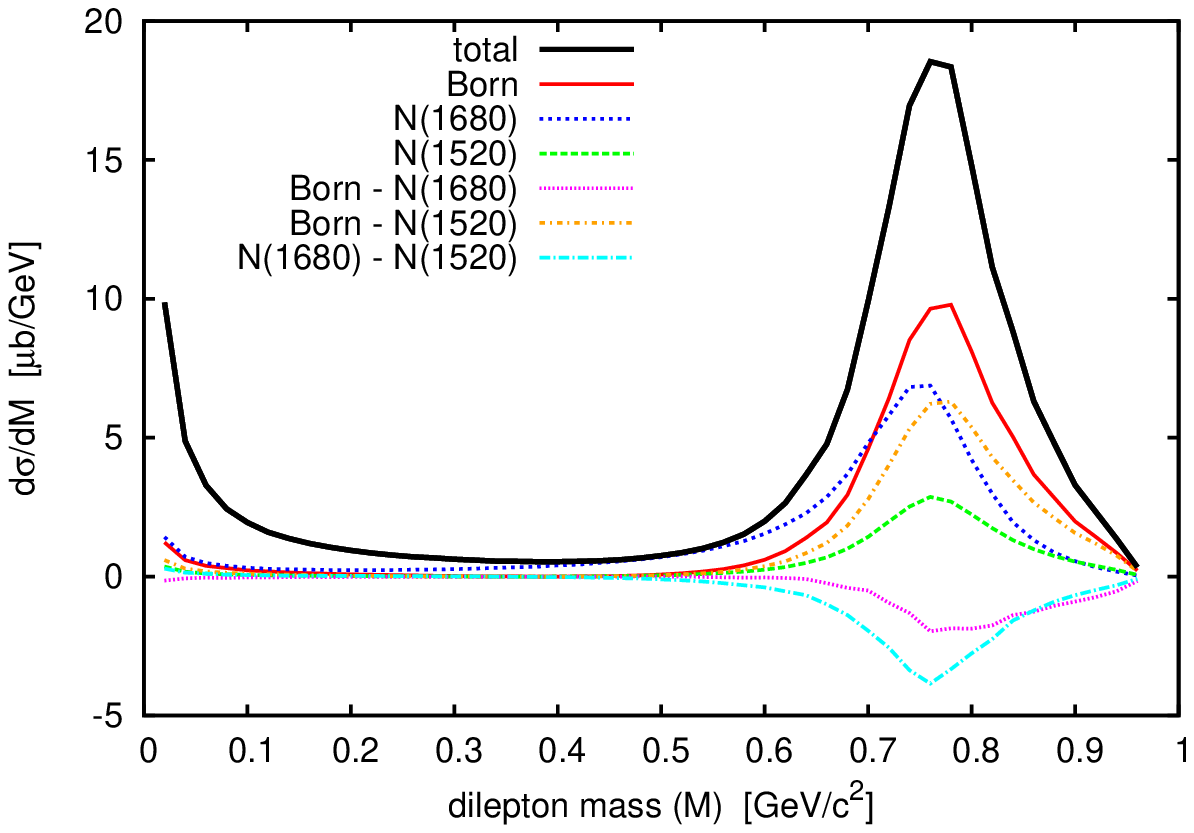}
  \caption{\label{fig:channels} (Color online) Contributions of the
    dominant channels to the dilepton invariant mass spectrum of the
    reaction $\pi^- + p \rightarrow n + e^+ + e^-$ at $\sqrt{s}=1.9$~GeV
    energy. See the text for the precise definition of the channels.}
  \end{center}
\end{figure}

As the center-of-mass energy increases from 1.3 GeV to 1.9 GeV the
importance of different resonances also changes. At 1.3 GeV the
$s$-channel $\Delta(1232)$ contribution dominates the dilepton cross
section. On the other hand at 1.9 GeV the Born term and the $s$-channel
$N(1680)$ gives the dominant contribution. The $s$-channel $N(1520)$
diagram is also important.  These can be seen in
Fig.~\ref{fig:channels} which shows the contributions of the dominant
channels to the dilepton spectrum at $\sqrt{s}=1.9$~GeV center-of-mass
energy.  Similarly to pion photoproduction, $u$-channel resonance
contributions are always negligible after the inclusion of the cutoff
Eq.~(\ref{eq:uch_cutoff}).

In Fig.~\ref{fig:channels} we also show the contribution of the
interference terms of the dominant channels. Note that interference
terms can be negative, therefore we used a linear scale on the vertical
axis. Since the interference terms are not negligible, dilepton
production in $\pi N$ collisions cannot be approximated by the
incoherent sum of $s$-channel baryon resonance diagrams
Fig.~\ref{fig:diagrams}(f), (which is the usual assumption in transport
models), even if a background term is added to simulate the Born term.
The simplest solution for transport models is to use the cross section
calculated by the sum of all diagrams shown in
Fig.~\ref{fig:diagrams}. There is a price to pay for that: it is
difficult to study in medium modification of baryon resonances in heavy
ion reactions.

\section{\label{sec:Conc} Conclusion}
We have developed an effective field theoretical model to calculate the
$\pi N \rightarrow Ne^+e^-$ cross section. We constructed an effective
Lagrangian including nucleons, photons, pions and $\rho$ mesons (via
VMD), and 16 baryon resonances below 2 GeV, i.e.\ all states with three-
or four-star status except $\Delta(1950)$ and $\Delta(1920)$.  We
applied form factors at each vertex for internal hadron lines to account
for their off-shell behavior.  To maintain gauge invariance we
generalized the method of Davidson-Workman \cite{Davidson-Workman} to
the production of massive photons (with and without an intermediate
$\rho$ meson). The $NN\pi$ and $\pi\pi\rho$ couplings are well known. In
the derivation of the interaction Lagrangians we used the
electromagnetic gauge invariance and a model inspired by SU(2) gauge
theory with $\rho$ mesons as gauge bosons. This model gives relations
between some of the coupling constants.

Coupling constants of baryon resonances to the $N\pi$ and $N\rho$
channels have been determined from the appropriate partial width of the
resonance, while the $RN\gamma$ couplings constants have been fitted to
the pion photoproduction data.

For dilepton production we obtained monotonically decreasing invariant
mass spectra below 1.5 GeV center-of-mass energy, while at higher
energies the VMD form factor (related to the intermediate $\rho$ meson)
creates a peak at high dilepton masses. The spectrum is dominated by the
Born-term, but the $N(1680)$ and $N(1520)$ and their interference terms
are sizable too. The importance of interference terms contradicts the
usual assumption of transport models that the cross section is dominated
by incoherent sum of $s$-channel resonance contributions.

\section*{ACKNOWLEDGMENTS}

The authors thank for the support by the Hungarian OTKA funds T71989 and T101438.
Gy.W. thanks support from the TET-10-1-2011-0061 and ZA-15/2009 joint
projects.

\end{document}